\documentclass[10pt]{article}
\usepackage{cite,epsfig,amssymb,amsmath,graphicx,color}

\allowdisplaybreaks

\evensidemargin \oddsidemargin

\begin{document}

\begin{titlepage}

\renewcommand{\thefootnote}{\fnsymbol{footnote}}


\vspace{15mm}
\baselineskip 9mm
\begin{center}
{\Large \bf Cylindrical symmetry locally isometric $AdS_{4}$ 
spacetime}
\end{center}

\baselineskip 6mm
\vspace{10mm}
\begin{center}
 Faizuddin Ahmed,$^1$ Davood Momeni,$^2$\footnote{Corresponding author} Debojit Sarma,$^1$ Bidyut Bikash Hazarika ,$^1$ Ratbay Myrzakulov $^2$
 \\[10mm]
  $^1${\sl Department of Physics, Cotton College Guwahati-781001, India}
  \\[3mm]
  $^2${\sl Eurasian International Center for Theoretical Physics and Department of General \& Theoretical Physics, Eurasian National University, Astana 010008, Kazakhstan}
     \\[10mm]
  {\tt faizuddinahmed15@gmail.com,~davoodmomeni78@gmail.com,\\~sarma.debojit@gamil.com,~bidyutbikash116@gmail.com,
rmyrzakulov@gmail.com}
\end{center}


\begin{center}
{\bf Abstract}
\end{center}
\noindent

We present a maximally symmetric vacuum spacetime, which is locally isometric anti- de Sitter, admitting closed timelike curves appear after a definite instant of time i.e., a time-machine spacetime. The spacetime is regular, free-from curvature singularities and an $4D$ extension of the Misner space in curved spacetime with identical causality violating properties.
\\ [13mm]


\end{titlepage}

\baselineskip 6.6mm
\renewcommand{\thefootnote}{\arabic{footnote}}
\setcounter{footnote}{0}


\newpage
\section{Introduction}
Anti-de Sitter space/Conformal Field Theory (AdS/CFT) conjecture
 has been a central paradigm for decades in theoretical 
physics from string theory to condensed matter physics \cite{Maldacena}. Underestanding and a fully description of the AdS space plays a crucial role in this paradaim. AdS spacetime is an exact solution for Einstein's field equations with a negative cosmological constant. The lowest-energy solution is the Ads spacetime and furthermore it was demonstrated that AdS is the unique static, asymptotically anti-de Sitter vacuum \cite{unique}.

Reciprocal transformations proposed by Buchdahl is a simple way to find exact solutions for Einstein field equations in the presene of a massless scalar field. \cite{Buchdahl}. Based on symmetry we can have both spherically or cylindrically metrics as far as we can have time dependent scalar field as sourceof the static metrics. Later Wynman proposed a simple series solution for the case of massless scalar field but when the scalar field is time ependent
\cite{Wyman}. A higher dimensional form of metric was found 
 in \cite{Xanthopolous}. Very recently 
 Vuille found an exact solution for Einstein gravity with 
a non zero (positive or negative)  cosmological constant in
 plane symmetry \cite{Vuille}.  Another interesting class of exact solutions are ones found by  Anzhong Wang with
  homothetic self-similarity \cite{Wang}. 
A natural generlization of the static solutions beyound four dimensions investigated by   Sarioglu  and Tekin
  \cite{Sarioglu}. In (2009) one of us found an exact solution for Einstein gravity in the presence of an arbitrary cosmological constant and with a static scalar field as a generalization of the Buchdahl extension of the Levi-Civita metric \cite{Miradhayeejafari:2009tu}. This solution is  a new cylindrically symmetric solution for
 a massless scalar field coupled minimally to gravity but contains
 cosmological constant with  regularity  for all  real values of scalar field parameter and with a  naked singularity for complex one

$AdS$ spacetimes could be obtain by embedding in a higher dimensional flat  space. This showed that for example $AdS_3$ were already embed ded into the $4D$ flat spacetime  $\mathcal{R}^{2,2}$ with metric $g=-(dX_1^2+dX_2^2)+(dX_3^2+dX_4^2)$. This embedding is invariant under $SO(2, 1)$,which is the precise isometry group of $AdS$, with confor- mal boundary. An appropriate rescaling casts the boundary in hyperboloid form in $(2)D$ form, which is universal for systems supporting $SO(1, 2)$ isometry group, with one dilatation and two special conformal transformations.\par
Some solutions of Einstein’s field equations have peculiar properties called closed causal curves (CCCs). These CCCs are closed timelike curves (CTCs), closed timelike geodesics (CTGs), closed null curves (CNC) or geodesics (CNGs) etc.. In literature, there are a number of solutions of field equations that have CCCs. A sample of these are \cite{Go}-\cite{Sar1}. One way of classifying such causality violating solutions would be to categorize the spacetime as either eternal time-machine in which CTCs always exist,(e.g. \cite{Go,Sto,Tip}. The second one being the true time-machine spacetime, where CTCs appears after a certain instant of time. In this direction Amos Ori deserves the special mention \cite{Ori}-\cite{Ori4}. We note that the eternal time-machine spacetimes are unrealistic models for a putative time- machine. A workable model of a time-machine must be a spacetime where CTCs appear at a definite instant of time. Among the true time-machine spacetime, we mention two: the first being the Oris vacuum core which is represented by a vacuum metric locally isometric to plane wave \cite{Ori2} and second, the Misner space in 2D \cite{Misn}. The Misner space is interesting in the context of CTCs as it is a prime example of a spacetime where CTCs evolve from causally well-behaved initial conditions.

The metric for the Misner space in 2D \cite{Misn}
\begin{equation}
ds^2=-2\,dt\,dx-t\,dx^2 \quad .
\label{1}
\end{equation}
where $-\infty < t < \infty$ but the co-ordinate $x$ periodic. The metric (\ref{1}) is regular everywhere as $\det g=-1$ including at $t = 0$. The curves $t = t_0$, where $t_0$ is a constant, are closed since $x$ is periodic. The curves $t < 0$ are spacelike, but $t > 0$ are timelike and the null curves $t=t_0=0$ form the chronology horizon. The second type of curves, namely, $t=t_0>0$ are closed timelike curves (CTCs).

The Einstein Field equation in electrovacuum is given by
\begin{equation}
G_{\mu\nu}+\Lambda\,g_{\mu\nu}=0\quad or \quad R_{\mu\nu}-\frac{1}{2}\,g_{\mu\nu}\,R+\Lambda\,g_{\mu\nu}=0\quad
\label{2}
\end{equation}
where $\mu,\nu\in(1,2,3,4)$. 
For conformally flat spacetime the Weyl tensor defined by
\begin{equation*}
C_{\alpha\beta\mu\nu}=R_{\alpha\beta\mu\nu}+\frac{1}{2}\Big(%
-R_{\alpha\mu}g_{\beta\nu}+R_{\alpha\nu}g_{\beta\mu}+R_{\beta\mu}g_{\alpha%
\nu}-R_{\beta\nu}g_{\alpha\mu}+\frac{1}{3}(g_{\alpha\mu}g_{\beta\nu}-g_{%
\alpha\nu}g_{\beta\mu})R)\Big)
\end{equation*}
vanishes, the Riemann tensor $R_{\mu\nu\rho\sigma}$ can be expressed in terms of metric tensor $g_{\mu\nu}$ by
\begin{equation}
R_{\mu\nu\rho\sigma}=k\,\left(g_{\mu\rho}\,g_{\nu\,\sigma}-g_{\mu\sigma}\,g_{\nu\,\rho}\right)\quad.
\label{4}
\end{equation}
$AdS_4$ space has simply the non trivial toplogy $\mathcal{R} \times \mathcal{H}^3$  is the hyperbolic metric in polar coordinates. Another interesting dual spacetime example is the so called AdS soliton, which its topology is $\mathcal{R} \times\mathcal{B}\times\mathcal{ T }$.

In this letter, we attempted to write down an axially symmetric form of anti-de Sitter space where closed timelike curves (CTCs) appear after a certain instant of time–a time-machine spacetime. The spacetime is an extension of four-dimensional Misner space and the time- machine behaviour is carried over to the AdS space. We note that there has been earlier attempt at constructing a Misner-like AdS spacetime by Li\cite{Li} having the line element
\begin{equation}
ds^2=-\left(dt-t\,d\chi\right)^2+\alpha^2\,d\chi^2+\left(dy-y\,d\chi\right)^2+\left(dz-z\,d\chi\right)^2
\label{5}
\end{equation}

In (\ref{5}), there are no CTCs when $t^2<\alpha^2+y^2+z^2$, but CTCs appear in the region with $t^2>\alpha^2+y^2+z^2$. However, this spacetime, as in the case of the Misner space, is multiply- connected, and we attempt to overcome this problem by constructing a Misner-like,  $AdS_4$ spacetime which is simply-connected.
\section{Review of exact solutions with cosmological constant in cylindrical coordinates}
We begin with a general cylindrically symmetric metric
in Weyl coordinates   $(t,r,\varphi,z)$,in natural units
\begin{eqnarray}
ds^{2}=-e^{u(r)}dt^2+dr^2+e^{v(r)}d\varphi^2+e^{w(r)}dz^2
\end{eqnarray}
Field equation for a massless minimally coupled scalar field in the
presence of a cosmological constant term $\Lambda$ is reading as:
\begin{eqnarray}
R_{\mu\nu}-\Lambda g_{\mu\nu}=\phi_{;\mu}\phi_{;\nu}
\end{eqnarray}
Metric functions are given by  $u_{i}=\{u(r),v(r),w(r)\},\phi\equiv \phi(r)$ , $\acute{f}=\frac{d f}{d
 r}$.
 The system of the field equations  are given by the following:
\begin{eqnarray}
2u_{i}''+u_{i}'\sum^{3}_{j=1}u_{j}'-4 \Lambda=0, i=\{1,2,3\}\\
2\sum^{3}_{j=1}u_{j}''+\sum^{3}_{j=1}u_{j}'^{2}-4 \Lambda=4\phi'^{2}
\end{eqnarray}
It was showed that the system has different exact solutions as follows:
\begin{itemize}

\item solution with $u(r)=constant,w(r)=constant$,$\phi'=0$
 
The metric reads:
\begin{eqnarray}\nonumber
ds^{2}=-dt^2+dr^2+(a r)^2d\varphi^2+dz^2
\end{eqnarray}
 By applying change of coordinates:
\begin{eqnarray}\nonumber
\tilde{\varphi}=a\varphi
\end{eqnarray}
we can find
\begin{eqnarray}\nonumber
ds^{2}=-dt^2+dr^2+r^2d\tilde{\varphi}^2+dz^2
\end{eqnarray}
Which obviously is flat space (locally)in cylindrical Weyl
coordinates and the conical parameter $a$
related to the gravitational mass per unit length of the spacetime,
$\eta$ \cite{plb} 
\begin{eqnarray}\nonumber
a=1-4\eta, \ \ 0<a<1.
\end{eqnarray}

\item solution with $u(r)=v(r)=w(r)\neq constant$,$\phi'\neq0$
 
In this case we obtain the following exact solution:
\begin{eqnarray}
ds^{2}=dr^2+e^{-2\sqrt{\frac{\Lambda}{3}}r}(\xi^2e^{2\sqrt{3\Lambda}r}+1)^{2/3}(-dt^2+d\varphi^2+dz^2)
\end{eqnarray}
It was reported in \cite{Miradhayeejafari:2009tu} as an exact
solution which contains two parameter, Cosmological constant 
$\Lambda$ and scalar field susceptibility parameter $\xi$ , permits one
time-like Killing fields $ \frac{\partial}{\partial t}$,  three space-like
Killing fields , which span a Euclidean group  $G(3)$ (see  \cite{Wang} for cosmological forms and see \cite{Momeni:2009tk}-\cite{Momeni:2015aea}   for extensions in modified gravity theories). 
\end{itemize}

By computing the Kretschmann scalar we find:
\begin{itemize}
 \item If $\xi^2>0$ i.e. the scalar field's parameter is a real constant ,
 no naked singularity at
 $r=r_{0}$.
\item 
 If   $\xi\in\mathcal{C}$ has  naked singularity 
\end{itemize}
This solution 
recoves safely 
  $LC\Lambda$   and Buchdahl
 solutions, 
in the limit of  the solution with $\Lambda\neq 0$,$\phi=constant$ ($LC\Lambda$ family):

\begin{eqnarray}
ds^{2}=dr^2+e^{\pm2\sqrt{\frac{\Lambda}{3}}r}(-dt^2+d\varphi^2+dz^2)
\end{eqnarray}
is a member  of general $LC\Lambda$ family and in   limit  $\Lambda=0$,$\phi\neq constant$ reduces to the Buchdahl family solution:
\begin{eqnarray}
ds^2=dr^2+[\frac{3}{2}(c_{1}r+c_{2})]^{\frac{2}{3}}(-dt^2+d\varphi^2+dz^2)
\end{eqnarray}

\section{Analysis of the spacetime}
Cylindrical symmetric metric
\begin{equation}
ds^2=dr^2+e^{2\,\alpha\,r}\,\left(dz^2-\beta\,t\,r\,d\phi^2-\beta\,r\,dt\,d\phi-\beta\,t\,dr\,d\phi \right)\quad
\label{6}
\end{equation}
where $\phi$ co-ordinate is periodic $0\leq\phi\leq2\,\pi$, and $\alpha$ is an integer and $\beta$ is real. We have used co-ordinates $x^0=t$, $x^1=r$, $x^2=\phi$ and $x^3=z$. The ranges of the other co-ordinates are $t,z \in (-\infty,\infty)$ and $0 \leq r < \infty$. The metric has signature $(-,+,+,+)$ and the determinant of the corresponding metric tensor $g_{\mu\nu}$, $\det\;g=-\frac{1}{4}\,r^2\,e^{6\,\alpha\,r}$, vanishes at $r=0$. We note that for spacetime (\ref{6}), Ricci scalar $R=-12\,\alpha^2$ and $k=-\alpha^2$. The Einstein tensor is the diagonal form given by
\begin{equation}
G^{\mu}_{\mu}=3\,\alpha^2=-\Lambda\,\delta^{\mu}_{\mu}\quad.
\label{7}
\end{equation}

Consider closed orbits of constant $t=t_0$, $r=r_0$ and $z=z_0$, the line element (\ref{6}) reduces to $(1)D$
form
\begin{equation}
ds^2=-\beta\,t\,r\,d\phi^2\quad
\label{8}
\end{equation}
These orbits are null curves for $t=t_0=0$, spacelike throughout for $t=t_0<0$, but become timelike for 
$t=t_0>0$, which indicates the presence of CTCs. Here we choose $r=r_0>0$, a constant and $\beta>0$ such that the above curves are timelike. Hence CTCs form at a specific instant of time satisfying $t=t_0>0$. 

It isIt is crucial to have analysis that the above CTCs evolve from an spacelike t = constant hypersurface and thus t is a time coordinate \cite{Ori2}. This can be ascertained by calculating the norm of the vector  $\nabla_{\mu} t$ or by determining the sign of the component $g^{tt}$ in the metric tensor $g^{\mu\nu}$ \cite{Ori2}. We find from metric (\ref{6}) that
\begin{equation}
g^{tt}=\frac{t\,\left(4\,r\,e^{-2\,\alpha\,r}+\beta\,t\right)}{\beta\,r^2}\quad.
\label{9}
\end{equation}

A hypersurface $t=\mbox{constant}$ is spacelike provided $g^{tt}<0$ for $t=t_0<0$, but become timelike for $t=t_0>0$ provided $g^{tt}>0$. The hypersurface $t=\mbox{constant}$ have a true singularity at $r=0$. Our analysis throughout the paper is restricted to the range $r>0$. Thus we consider $r=r_0>0$, a constant and $\beta$ is sufficiently small positive number (ensuring that the bracket term (\ref{9}) is positive for $t<0$) such that $t=\mbox{constant}$<0 hypersurface is spacelike. Thus $t=\mbox{constant}$ spacelike hypersurface can be choosen as initial conditions over which the initial data may be specified. There is a Cauchy horizon for $t = t_0 = 0 $ called chronology horizon which separates the causal and non-causal of the spacetime. Hence the spacetime evolves from a Partial Cauchy hypersurface in a causally well-behaved manner, upto a moment, i.e., a null hypersurface $t = t_0 = 0$ and CTCs evolve from this spacelike Cauchy hypersurface at a specific instant of time.

\section{Axially symmetry of the spacetime}

Consider the following Killing vector $\eta=\partial_{\phi}$ which has the normal form 
\begin{equation}
\eta^{\mu}=\left (0,0,1,0\right )\quad. 
\label{10}
\end{equation}
It's co-vector form given by 
\begin{equation}
\eta_{\mu}=-\beta\,e^{2\,\alpha\,r}\,\left(\frac{r}{2},\frac{t}{2},r\,t,0\right )\quad
\label{11} 
\end{equation}  
Killing vector (\ref{10}) satisfies the Killing equation $\eta_{\mu\,;\,\nu}+\eta_{\nu\,;\,\mu}=0$. For cyclically symmetric metric, the norm $\eta_{\mu}\,\eta^{\mu}$ of the Killing vector is spacelike, closed orbits. We note that 
\begin{equation}
\eta^{\mu}\,\eta_{\mu}=-\beta\,t\,r\,e^{2\,\alpha\,r}\quad,
\label{12}
\end{equation}
which is spacelike for $t<0$, closed orbits (as $\beta$ is positive real nubmer). For axially symmetric, the norm (\ref{12}) must vanishes on the axis $r=0$ \cite{Marc,Wang,Wang2,Steph} is satisfied. The regularity-condition, namely that in the limit of rotation axis  $(r\rightarrow 0)$ one must have \cite{Marc,Wang,Wang2,Steph}
\begin{equation}
\frac{(\nabla_a(\eta_{\mu}\eta^{\mu}))(\nabla^a(\eta_{\mu}\eta^{\mu}))}{4\,\eta_{\mu}\eta^{\mu}}\rightarrow 1
\label{13}
\end{equation}
holds for the metric defined by (\ref{6}) for $t<0$. This ensures the validity of imposing the $2\,\pi$ periodicity on the $\phi$ co-ordinate. Hence the spacetime (\ref{6}) is regular on the axis and  cylindrially symmetric with $r$ is the true radial coordinate.

In spacetime (\ref{6}), we have done transformation $t\rightarrow \frac{t}{\beta\,r}$
\begin{equation}
ds^2=dr^2+e^{2\,\alpha\,r}\,\left(dz^2-t\,d\phi^2-dt\,d\phi\right)\quad
\label{15}
\end{equation}
Still a co-ordinate singularity $t\rightarrow \pm \infty$. To remove this singularity, we have done transformation 
\begin{equation}
\phi\rightarrow \phi-\frac{1}{2}\,\ln {t}\quad
\label{16}
\end{equation}
followed by
\begin{eqnarray}
t\rightarrow \left(\phi^2-t^2\right)\quad\quad \mbox{and} \quad\quad \phi\rightarrow \tanh^{-1}(\frac{\phi}{t})\quad
\label{17}
\end{eqnarray}
We get the diagonal form of the metric  (\ref{15})
\begin{equation}
ds^2=dr^2+e^{2\,\alpha\,r}\,\left(dz^2+d\phi^2-dt^2\right)\quad 
\label{18}
\end{equation}

Substituting $\alpha=\pm\,\sqrt{-\frac{\Lambda}{3}}$ using (\ref{7})  into (\ref{18}), we get the form \cite{Zof},\cite{Miradhayeejafari:2009tu}
\begin{equation}
ds^2=dr^2+e^{\pm\,2\,\sqrt{-\frac{\Lambda}{3}}\,r}\,\left(dz^2+d\phi^2-dt^2 \right)\quad 
\label{19}
\end{equation}
Spacetime with metric (\ref{19}) constitutes a part of the anti-de Sitter spacetime with the metric written in the horospherical coordinates. This follows from the following transformation into the metric  (\ref{19})
\begin{eqnarray}
\nonumber
r&=&\mp\,\sqrt{-\frac{3}{\Lambda}}\,\ln (\sqrt{-\frac{\Lambda}{3}}\,x)
\end{eqnarray}
One get conformally flat standard form of locally isometric  $AdS_4$ space \cite{Zof}
\begin{equation}
ds^2=\frac{3}{(-\Lambda)\,x^2}\,\left(-dt^2+dx^2+d\phi^2+dz^2\right)\quad
\label{21}
\end{equation}
where $\Lambda<0$ and one of the co-ordinate $\phi$ being periodic. Solution (\ref{21}) with the `$+$' sign covers that part of the anti-de Sitter hyperboloid where $x\in[\sqrt{\frac{3}{(-\Lambda)}},\infty)$ while the `$-$' branch is valid for $x\in[\sqrt{\frac{3}{(-\Lambda)}},0)$. However, metric (\ref{6}) is axially symmetric and $r=0$ represents the axis of rotation. The spacetime (\ref{6}) is free-from curvature singularities and the curvature invariant
(Kretchsmann scalar K)
$$K=R_{\mu\nu\rho\sigma}\,R^{\mu\nu\rho\sigma}=24\,\alpha^2$$ which is a constant.

Consider a null vector $k_{\mu}$ for metric (\ref{6}) whose normal form given by
\begin{equation}
k_{\mu}=\left(0,0,\frac{1}{2},0\right)\quad
\label{n1}
\end{equation}
The null vector (\ref{n1}) satisfy the geodesics equation
\begin{equation}
k_{\mu\,;\,\nu}\,k^{\nu}=0\quad
\label{24}
\end{equation}
Using the null vector (\ref{n1}) we have calculated the {\it Optical} scalar \cite{Steph}, the {\it expansion}, the {\it twist} (rotation) and the {\it shear} and they are
\begin{eqnarray}
\boldsymbol{\theta}&=&\frac{1}{2}\,k^{\mu}_{\,;\,\mu}=0\\
\boldsymbol{\omega^2}&=&\frac{1}{2}\,{k_{[\mu\,;\,\nu]}}\,k^{\mu\,;\,\nu}=0\\
\boldsymbol{\sigma}\,\boldsymbol{\bar{\sigma}}&=&\frac{1}{2}\,{k_{(\mu\,;\,\nu)}}\,k^{\mu\,;\,\nu}-\boldsymbol{\theta^2}=0\quad.
\end{eqnarray}
Hence the spacetime (\ref{6}) is  is non-diverging, have shear-free null geodesics congruence.
\section{Fermi-Robertson-Walker (FRW) flat space}

It is interesting to note that by doing a compolex transformation into the spacetime (\ref{19})
\begin{equation}
r\rightarrow i\,\tau\quad and\quad t\rightarrow i\,u\quad
\label{25}
\end{equation}
one get 4D de-Sitter spacetime with $\Lambda>0$ as 
\begin{equation}
ds^2=-d\tau^2+e^{\mp\,2\,\sqrt{\frac{\Lambda}{3}}\,\tau}\,\left(d\phi^2+dz^2+du^2 \right)\quad.
\label{26}
\end{equation}
The spacetime (\ref{26}) can be written as
\begin{equation}
ds^2=-d\tau^2+{f^2(\tau)}\,\left(d\phi^2+dz^2+du^2\right)\quad,
\label{27}
\end{equation}
where 
\begin{equation}
f(\tau)=e^{\mp\,H_0\,\tau}\quad \mbox{and} \quad H_0=\sqrt{\frac{\Lambda}{3}}\quad.
\label{28}
\end{equation}
The spacetime (\ref{27}) with (\ref{28}) is the Fermi-Robertson-Walker (FRW) cosmological model in $(\tau,u,\phi,z)$. The metric (\ref{27}) is increasing with increasing time due to exponential factor in case of `$+$', hence the model is used as expanding universe. For expanding universe, transforming $t_e \rightarrow -H_0^{-1}\,e^{-H_0\,t}$ into the metric (\ref{27}) one get
\begin{equation}
ds^2=\frac{1}{H_0^2\,t_e^{2}}\,\left(-dt_e^{2}+du^2+d\phi^2+dz^2\right)\quad,
\label{29}
\end{equation}
where $t_e$ runs from $-\infty$ in the far past to $0$ in the far future and the Hubble parameter is given by $H=\frac{\dot{f}}{f}=H_0$\quad.

While the model (\ref{27}) is used as contracting universe in case of `$-$' sign as time increases. Hence for contracting universe, transforming $t_c\rightarrow H_0^{-1}\,e^{H_0\,t}$ into the metric (\ref{27}) one get
\begin{equation}
ds^2=\frac{1}{H_0^2\,t_c^{2}}\,\left(-dt_c^{2}+du^2+d\phi^2+dz^2\right)\quad,
\label{30}
\end{equation}
where $t_c\in[0,\infty]$ and the Hubble parameter is given by $H=\frac{\dot{f}}{f}=-H_0$\quad.

\section{Conclusion}
We have presented cylindrical symmetric model, an extension of the Misner space in curved spacetime, admitting closed timelike curves. These curves evolves from a cauchy spacelike hypersurface after a certian instant of time and hence can be used as Time-Machine model. Our primary motivation in this paper is to write down a metric for a spacetime that incorporates the Misner space and its causality violating properties and to classify it. The solution is an axially symmetric metric (5) which serves as a model of a Time-Machine spacetime in the sense that CTCs appear at a definite instant. We also note that the spacetime represented  contains the symmetry axis rendering it simply-connected. In contrast, the Misner space is multiply-connected.




\begin{thebibliography}{99}
\bibitem{Maldacena}
J. Maldacena , Adv. Theor. Math. Phys. 2, 231(1999)[ Int. J. Theor. Phys. 38, 1113 (1999)].
\bibitem{unique}
W. Boucher, G. W. Gibbons, and Gary T. Horowitz,Phys. Rev. D (3) 30 (1984), no. 12, 2447–2451. MR 86e:83014.

\bibitem{Buchdahl}
H. A. Buchdahl. Phys. Rev. 115, 1325-1328 (1959)

\bibitem{Momeni:2009tk} 
D.~Momeni and H.~Gholizade,
  Int.\ J.\ Mod.\ Phys.\ D {\bf 18}, 1719 (2009)
  doi:10.1142/S0218271809015266
  [arXiv:0903.0067 [gr-qc]].
\bibitem{Wyman}
M. Wyman. Phys. Rev. D. vol 40, Num  8 (1981)
\bibitem{Xanthopolous}
B. C. Xanthopolous and T. Zannias. Phys. Rev. D. vol 40, Num 8
(1989)

\bibitem{Vuille}
C.  Vuille. Gen Relativ Gravit (2007) 39:621ñ632 (2007)

\bibitem{Wang}
A. Z. Wang. Phys. Rev. D.68 (2003) 064006; Anzhong Wang : Phys.Rev. D72 (2005) 108501.

 \bibitem{Sarioglu}
  O.~Sarioglu and B.~Tekin,
  Phys.\ Rev.\ D {\bf 79}, 087502 (2009)
  doi:10.1103/PhysRevD.79.087502
  [arXiv:0901.1242 [gr-qc]].
\bibitem{Miradhayeejafari:2009tu}
   D. Momeni and H. Miraghaei,
  Int.\ J.\ Mod.\ Phys.\ A {\bf 24} (2009) 5991
  doi:10.1142/S0217751X09046369
  [arXiv:0903.5171 [gr-qc]].
\bibitem{plb}
  A.~Azadi, D.~Momeni and M.~Nouri-Zonoz,
  Phys.\ Lett.\ B {\bf 670}, 210 (2008)
  doi:10.1016/j.physletb.2008.10.054
  [arXiv:0810.4673 [gr-qc]].
\bibitem{Momeni:2009tk} 
  D.~Momeni and H.~Gholizade,
ote on constant curvature solutions in cylindrically symmetric metric $f(R)$ Gravity,''
  Int.\ J.\ Mod.\ Phys.\ D {\bf 18}, 1719 (2009)
  doi:10.1142/S0218271809015266
  [arXiv:0903.0067 [gr-qc]].
\bibitem{Momeni:2009au} 
  D.~Momeni,
  Int.\ J.\ Theor.\ Phys.\  {\bf 50}, 1493 (2011)
  doi:10.1007/s10773-010-0659-9
  [arXiv:0910.0594 [gr-qc]].

\bibitem{Rodrigues:2012qu} 
  M.~E.~Rodrigues, M.~J.~S.~Houndjo, D.~Momeni and R.~Myrzakulov,
  Can.\ J.\ Phys.\  {\bf 92}, 173 (2014)
  doi:10.1139/cjp-2013-0414
  [arXiv:1212.4488 [gr-qc]].

\bibitem{Houndjo:2013us} 
  M.~J.~S.~Houndjo, M.~E.~Rodrigues, D.~Momeni and R.~Myrzakulov,
  Can.\ J.\ Phys.\  {\bf 92}, no. 12, 1528 (2014)
  doi:10.1139/cjp-2014-0070
  [arXiv:1301.4642 [gr-qc]].

\bibitem{Bravetti:2012hd} 
  A.~Bravetti, D.~Momeni, R.~Myrzakulov and H.~Quevedo,
  Gen.\ Rel.\ Grav.\  {\bf 45}, 1603 (2013)
  doi:10.1007/s10714-013-1549-2
  [arXiv:1211.7134 [gr-qc]].

\bibitem{Momeni:2010jf} 
  D.~Momeni, M.~R.~Setare and N.~Majd,
  JHEP {\bf 1105}, 118 (2011)
  doi:10.1007/JHEP05(2011)118
  [arXiv:1003.0376 [hep-th]].


\bibitem{Sebastiani:2013fsa} 
  L.~Sebastiani, D.~Momeni, R.~Myrzakulov and S.~D.~Odintsov,
  Phys.\ Rev.\ D {\bf 88}, no. 10, 104022 (2013)
  doi:10.1103/PhysRevD.88.104022
  [arXiv:1305.4231 [gr-qc]].



\bibitem{Momeni:2015aea} 
  D.~Momeni, K.~Myrzakulov, R.~Myrzakulov and M.~Raza,
  arXiv:1505.08034 [gr-qc].


\bibitem{Go} K. G\"{o}del, Rev. Mod. Phy. {\bf 21}, 447 (1949). 
\bibitem{Sto} W. J. van Stockum, Proc. R. Soc. Edin. {\bf 57}, 135 (1937).
\bibitem{Tip} F. J. Tipler, Phys. Rev. {\bf D9}, 2203 (1974).
\bibitem{MTY} M. S. Morris, K. S. Thorne and U. Yurtsever, Phys. Rev. {\bf D49}, 3990 (1988) ;  Phys. Rev. Lett. {\bf 61}, 1446 (1988).
\bibitem{Gott} J. R. Gott, Phys. Rev. Lett. {\bf 66}, 1126 (1991).
bibitem{Ste} B. R. Steadman, Class. Quantum Grav. {\bf 35}, 1721 (2003).
\bibitem{Bon} W. B. Bonnor and B. R. Steadman, Gen. Rel. Grav. {\bf 37}, 1833 (2005).
\bibitem{Taub} C. W. Misner and A. H. Taub, Sov. Phys. JETP, {\bf 28}, 122 (1969).
\bibitem{Kerr} R. P. Kerr, Phys. Rev. Lett. {\bf 11}, 237 (1963).
\bibitem{Cart} B. Carter, Phys. Rev. {\bf 174}, 1559 (1968).
\bibitem{Haw} S. W. Hawking and G. F. R. Ellis, {\it The Large Scale Structure of Spacetime}, Cambridge University Press, (1973).
\bibitem{Gron} \O. Gr{\o}n and S. Johannesen, New Journal of Physics, {\bf 10}, 103025 (2008).
\bibitem{Gron2} \O. Gr{\o}n and S. Johannesen, Nuovo Cimento, {\bf B125}, 1215 (2010).
\bibitem{Sar1} D. Sarma, M. Patgiri and F. U. Ahmed, Ann. of Phys. {\bf 329}, 184 (2013).
\bibitem{Ori} A. Ori, Phys. Rev. {\bf D76}, 044002 (2007).
\bibitem{Ori2} A. Ori, Phys. Rev. Lett. {\bf 95}, 021101 (2005).
\bibitem{Misn} C. W. Misner, {\it Relativity Theory and Astrophysics I : Relativity and Cosmology}, edited by J. Ehlers, {\it Lectures in Applied Mathematics},{\bf Vol.8}, AMS (1967).

\bibitem{Ori3} A. Ori, Phys. Rev. {\bf D44}, R2214 (1991).
\bibitem{Ori4} A. Ori, Phys. Rev. Lett. {\bf 71}, 2517 (1993).
\bibitem{Li} L.-X. Li, Phys. Rev. {\bf D59}, 084016 (1999).
\bibitem{Marc} M. Mars and J. M. M. Senovilla, Class. Quantum Grav. {\bf 10}, 1633(1993) ; Class. Quantum Grav. {\bf 12}, 2071 (1995) ; Class. Quantum Grav. {\bf 20}, L293-L300 (2003).
\bibitem{Wang} J. C. N. de Araujo and Anzhong Wang, Gen. Rel. Grav. {\bf 32}, 1971 (2000).
\bibitem{Wang2} Anzhong Wang, Phys. Rev. {\bf D68}, 064006 (2003).
\bibitem{Steph} H. Stephani, D. Kramer, M. MacCallum, C. Hoenselaers, E. Herlt, {\it Exact Solutions to Einstein's Field Equations},Cambridge Univ. Press, Cambridge,(2003).
\bibitem{Zof} M. \v{Z}ofka, J. Bi\v{c}\'{a}k, Class. Quantum Grav. {\bf 25}, 015011 (2008).


  
\end{thebibliography}
\end{document}